# Unconventional Excitonic States with Phonon Sidebands in Layered Silicon Diphosphide

Ling Zhou[1†], Junwei Huang[1†], Lukas Windgaetter[2†], Chin Shen Ong[3], Xiaoxu Zhao[4], Caorong Zhang[1], Ming Tang[1], Zeya Li[1], Caiyu Qiu[1], Simone Latini[2], Yangfan Lu[5,10], Di Wu[1], Huiyang Gou[6], Andrew T. S. Wee[7], Hideo Hosono[5], Steven G. Louie[3], Peizhe Tang[8,2*], Angel Rubio[2,9*], Hongtao Yuan[1*]

[1]*National Laboratory of Solid State Microstructures, Jiangsu Key Laboratory of Artificial Functional Materials and College of Engineering and Applied Sciences, Nanjing University, Nanjing 210000, China.*

[2]*Max Planck Institute for the Structure and Dynamics of Matter, Center for Free Electron Laser Science, 22761 Hamburg, Germany.*

[3]*Department of Physics, University of California at Berkeley, and Materials Sciences Division, Lawrence Berkeley National Laboratory, Berkeley, California 94720, USA.*

[4]*School of Materials Science and Engineering, Nanyang Technological University, 637371, Singapore.*

[5]*Materials Research Center for Element Strategy, Tokyo Institute of Technology, 4259 Nagatsuta, Midori-ku, Yokohama 226-8503, Japan.*

[6]*Center for High Pressure Science and Technology Advanced Research, Beijing 100094, China.*

[7]*Department of Physics, National University of Singapore, 2 Science Drive 3, 117542, Singapore.*

[8]*School of Materials Science and Engineering, Beihang University, Beijing 100191, China.*

[9]*Center for Computational Quantum Physics, Simons Foundation Flatiron Institute, New York, NY 10010 USA.*

[10]*College of Materials Science and Engineering, National Engineering Research Center for Magnesium Alloys, Chongqing University, Chongqing 400030, China*

E-mail: htyuan@nju.edu.cn; peizhet@buaa.edu.cn; angel.rubio@mpsd.mpg.de

[†] These authors contributed equally to this work.

**Many-body interactions between quasiparticles (electrons, excitons, and phonons) have led to the emergence of new complex correlated states and are at the core of condensed matter physics and material science. In low-dimensional materials, unique electronic properties for these correlated states could significantly affect their optical properties. Herein, combining photoluminescence, optical reflection measurements and theoretical calculations, we demonstrate an unconventional excitonic state and its bound phonon sideband in layered silicon diphosphide ($SiP_2$), in which the bound electron-hole pair is composed of electrons confined within one-dimensional phosphorus–phosphorus chains and holes extended in two-dimensional $SiP_2$ layers. The excitonic state and the emergent phonon sideband show linear dichroism and large energy redshifts with increasing temperature. Within the *GW* plus Bethe–Salpeter equation calculations and solving the generalized Holstein model non-perturbatively, we confirm that the observed sideband feature results from the correlated interaction between excitons and optical phonons. Such a layered material provides a new platform to study excitonic physics and many-particle effects.**

An exciton, the electron–hole pair formed via Coulomb interaction, is an ideal platform for understanding many-body effects[1–8]. The properties of excitons strongly depend on the crystal structure and dimensionality of host materials[9–11]. Due to quantum confinement, the electronic properties of quasiparticles (electrons, holes and excitons) in low-dimensional materials can be remarkably different from those in three-dimensional (3D) bulk materials. The Coulomb screening in low-dimensional quantum-confined structures, particularly in one-dimensional (1D) electronic systems, is known to be weaker than that in bulk systems and consequently leads to larger exciton binding energy[12–14] and other emergent excitonic phenomena[10]. Experimental observations of anisotropic excitons have been demonstrated in two-dimensional (2D) van der Waals (vdWs) materials in which electrons and holes taking part in the formation of 2D excitons are confined in the same monolayer[15–19]. Meanwhile, in 1D materials such as carbon nanotubes (CNTs)[20,21], 1D excitonic states have also been observed, in which constituent electrons and holes are known to be confined within 1D nanostructures. A unique excitonic state with hybrid dimensionality, which is yet elusive, such as a bound electron-hole pair with an electron confined along one dimension (1D-confined electron) and a hole confined along two dimensions (2D-confined hole), or vice versa, would be of great interest and highly desired in terms of its optical properties and interactions with other emergent quasiparticles.

In this work, we demonstrate the observation of an unconventional bright exciton in a layered $SiP_2$ crystal, accompanied by a correlated phonon sideband in the optical spectrum. Based on our *ab initio* many-body *GW* and *GW* plus Bethe–Salpeter equation (*GW*–BSE) calculations, as well as non-perturbative model calculations, we find that the electrons constituting the excitons are confined within the 1D phosphorus–phosphorus chains of $SiP_2$, while the correlated holes extend over the 2D $SiP_2$ atomic plane. Therefore, excitonic states in layered $SiP_2$ are expected to exhibit hybrid dimensionality properties. Photoluminescence (PL) spectroscopy and reflectance contrast (RC) spectroscopy show that, regardless of the polarization of the excitation laser, the optical response of the excitonic state is always linearly polarized along the $x$ direction of the $SiP_2$ lattice and is accompanied by a unique sideband feature. Both the excitonic emission and the sideband feature undergo dramatic redshifts as the temperature increases, in contrast with a slight temperature-dependent redshift of the band edge that is mainly influenced by electron–phonon coupling[22,23]. This reveals that in $SiP_2$, the interaction of the electronic degrees of freedom with the phononic degrees of freedom is strongly enhanced by excitonic effects. The phonon sideband feature can be theoretically modeled using a non-perturbative approach to describe the interaction between the unconventional excitons and optical phonon modes. Note that normally, the reduced dimensionality leads to excitonic features that are strongly affected by extrinsic environmental effects, such as disorder from the substrate and surface additives[10]. Here, we provide an investigation on the intrinsic excitonic behavior in thicker, bulk-like $SiP_2$ flakes. Such a tightly bound unconventional exciton in $SiP_2$ not only can be envisioned as a new platform for the exploration of exciton–phonon (ex–ph) coupling[24–28] and other many-body physics but also may lend itself to potential applications for anisotropic optoelectronic devices.

Layered $SiP_2$ is chosen as our target material because of the following unique characteristics. Compared with hexagonal layered materials such as graphene and $MoS_2$, the cleavable $SiP_2$ crystal (space group *Pnma*) possesses an orthorhombic layered structure with a huge in-plane lattice anisotropy, as schematically shown in Fig. 1a and experimentally confirmed by scanning transmission electron microscopy–annular dark field (STEM–ADF) imaging in Fig. 1b–d and Figs. S1 and S2 in the Supplementary Information (SI). Remarkably, based on their atomic surroundings, two types of inequivalent phosphorus atoms $P_A$ and $P_B$ can be distinguished in the $SiP_2$ lattice. As shown in Fig. 1a, $P_A$ binds to three silicon (Si) atoms, while $P_B$ binds to one Si atom and the other two equivalent $P_B$ atoms. Note that the $P_B$ atoms along the $y$ direction of the crystal lattice can naturally form phosphorus–phosphorus chains (denoted as $P_B$–$P_B$ chains) embedded in the bulk $SiP_2$ (blue shades in Fig. 1a), which play a critical role in realizing the quasi-1D electronic states involved in exciton formation. To identify the variation in the chemical bonding environment around $P_A$ and $P_B$ atoms and the resulting unique properties of $P_B$–$P_B$ chains in layered $SiP_2$, we performed arsenic doping experiments (Sec. 2-4 of SI) and used STEM characterization (Fig. S5). One can see that the doped arsenic atoms only selectively substitute the $P_B$ atoms inside the $P_B$–$P_B$ chains (more details in Fig. S5), indicating that the atomic structure containing $P_B$–$P_B$ chains in $SiP_2$ is distinct from the buckled structure in black phosphorus (BP).

More importantly, the anisotropy induced by quasi-1D $P_B$–$P_B$ chains in layered $SiP_2$ directly results in unique electronic properties. Figure 1e shows the band structure of semiconducting bulk $SiP_2$ obtained from *GW* calculations. We found that the conduction band edge states in the X–Γ–Z plane of the first Brillouin zone (BZ) are relatively flat with a large effective mass (Table S1), and the corresponding charge densities are localized on the $P_B$–$P_B$ chains (Fig. 1f), behaving like 1D-confined electrons. Importantly, in the direction along the $P_B$–$P_B$ chains (the $y$ direction of the crystal lattice), the electron hopping on P$_B$ atoms is significantly larger (bandwidth ≈ 1.63 eV) than that across the P$_B$–P$_B$ chains (bandwidth ≈ 0.08 eV) (see details in Fig. S22), confirming the 1D nature of this electronic state on the conduction band edge. On the other hand, the hole states at the valence band edge do not show the same level of anisotropy (see Sec. 2-2 of SI), which, compared with 1D electrons, are relatively extended over the whole atomic plane in a quasi-2D fashion. The hybrid dimensionality of these band edge states in SiP$_2$ is remarkably different from those of the anisotropic 2D states in BP[15,29]. By analyzing the calculated phonon bands given in Sec. 13 of SI, we identify that the optical phonons localized on P$_B$ atoms and neighboring Si atoms could have a large coupling with quasi-1D electronic states in layered SiP$_2$.

Figure 2a–b presents the PL spectrum and the 2$^{nd}$ derivative of the RC (d-RC, see Sec. 6 in SI) of a SiP$_2$ flake (228 nm) at 5.5 K, which reflects the light emission and absorption properties, respectively. The PL spectrum shows a main peak A at 2.06 eV (the lowest bright excitonic bound state denoted as the A exciton) and a broadened sideband feature A′ at 2.01 eV. The main peak A, obtained from all SiP$_2$ flakes measured at 5.5 K, is consistently located at an emission energy of 2.06 ± 0.01 eV (here, 0.01 eV is the energy uncertainty obtained from the standard deviation of the emission energies of several measured SiP$_2$ flakes; see Sec. 7 of SI). Such a peak A in the PL spectrum matches the peak at 2.05 eV in the d-RC spectrum, as indicated by the red arrow (Fig. 2b). Due to the interference of the RC signals from the different interfaces in the SiP$_2$ thin films supported by substrates (see Sec. 6 of SI), the phonon sideband feature is difficult to identify from the d-RC spectrum.

Figure 2c shows the absorbance spectra obtained from the *GW*–BSE calculation and *GW* calculation with the random phase approximation (*GW*–RPA). Compared with the calculated absorption spectra based on *GW*–BSE and *GW*–RPA, we confirm that the emission peak A at 2.12 eV originates from the recombination of an excitonic state, in which the electronic states for electrons are quasi-1D and related electronic states for holes are quasi-2D (Fig. 1f). As shown in Fig. 2c–d, the calculated binding energy of such an unconventional exciton is approximately 140 meV (for more details about *GW*–BSE calculations, see the Methods Section and Sec. 12 of SI). From the modulus squared of the exciton wavefunction in real space shown in Fig. 3c, the observed exciton behaves like a Wannier-type exciton with twofold rotational symmetry, in sharp contrast to 2D excitons in monolayer transition-metal-dichalcogenides[13]. More importantly, this exciton is embedded in a bulk layered material with an unusual atomic structure in contrast to those reported pure 1D excitons in semiconductor nanowires[30] and CNTs[12,31], leading to strongly anisotropic Coulomb screening for 1D-confined electrons and 2D-confined holes.

Since the unconventional A exciton is mainly contributed by electrons and holes localized along the X–Γ–Z direction in the first BZ (Sec. 12-2 of SI), we use the band edge states at the X point as the representative **k**-point to explore the influence of electron–phonon interactions on its electronic structures and optical response. Figure 2f shows the zero-point energy shifts of the band gap at the X point induced by all optical phonon modes with momentum **q** = 0 at zero temperature. Here, we use the frozen-phonon approximation[32] to estimate the influence of optical phonon vibrations on the electronic states at the X point (see Sec. 13-2 of SI for more details). Since the electron wavefunctions of the A exciton are localized on the $P_B$–$P_B$ chains, this unconventional exciton couples most strongly to optical phonons, whose vibrational modes are in the X–Γ–Z plane and involve $P_B$ atoms and neighboring Si atoms (Fig. 2g). These optical phonon modes dramatically modify the electronic structures of the quasi-1D states (Sec. 13-4 of SI), indicating significant electron–phonon coupling within the $P_B$–$P_B$ chains. Comparing the results in Fig. 2f–g, one can see that the prominent energy shifts are from the optical phonon modes with eigenenergies of 50 ~ 60 meV. More details are given in Sec. 13-2 and 13-4 of SI.

The experimental observation of the sideband feature A′ also indicates that ex–ph interaction on the quasi-1D $P_B$–$P_B$ chains is at least moderately strong (see Sec. 8 of SI). Therefore, we use a non-perturbative model to simulate the emergence of the sideband feature A′, where a "generalized Holstein Hamiltonian" is used with inputs from first-principles calculations, and the self-energy effects are included beyond the first-order Fan–Migdal diagram (see Methods Section). In this model, we found that the fitted ex–ph coupling constant $M$ of 30 meV is comparable to the relatively small bandwidth (or hopping, $t_{ex}$ = –20 meV) of the unconventional A exciton (for the estimate of $t_{ex}$, see Methods Section and Sec. 13-3 of SI). Our approach is similar to the cumulant method considering the ex–ph coupling within the perturbative limit and makes use of the exponential assumption to include the self-energy effects from higher-order diagrammatic terms[33–35]. As shown in Fig. 2e, the appearance of the phonon sideband peak in the simulated spectrum agrees with the experimental results, indicating that sideband A′ originates from the ex–ph coupling between the unconventional exciton and the abovementioned optical phonon modes.

Figure 3a–b shows the contour plots of the PL and d-RC intensity of bulk SiP$_2$ as a function of emission energy at different detection polarization angles $\theta$ ($\theta = 0°$ is set along the *x* direction), suggesting that the linear dichroic absorption and PL emission have similar twofold symmetry characteristics (see Sec. 4 and Sec. 12 of SI for more details). Note that the observed linearly polarized PL emission remains along the *x* direction regardless of the incident laser polarization direction or the sample temperature, as shown in Figs. S10 and S11. Our *GW*–BSE calculations (Figs. 3d and S21b) show that the absorption peak of the quasi-1D A exciton appears only when the polarization is along the *x* direction. The absorption signal inside the band gap along the *y* direction is forbidden by the SiP$_2$ crystal symmetry, which results in relevant optical excitonic matrix elements being zero (Sec. 12-3 of SI). We also performed pump-probe transient optical measurements to characterize the dynamics of the observed bright exciton in bulk SiP$_2$ (see Sec. 9 of SI for more details). The lifetime for the exciton in SiP$_2$ is as short as 250 fs, which is probably related to an ultrafast process that dissociates these linearly polarized bound excitonic states into unbound and unpolarized states.

We further compare the energy shift of the A exciton peak with the energy shift of the quasiparticle band edge as the temperature changes. Figure 4a–b shows the temperature-dependent PL and d-RC spectra for bulk SiP$_2$. As shown in Fig. 4c, the optical absorption of the band edges and the exciton peak A, as well as the sideband feature A′, all exhibit clear redshifts with increasing temperature. The redshifts of the band edge can be fitted with the Bose–Einstein model (see Sec. 7 of SI for more details), suggesting that the interaction between electrons and phonons plays an important role in the energy shifts. The redshift of the band edge absorption resulting from the electron–phonon coupling[22,23,36] is approximately 20 meV at 300 K. On the other hand, the redshift of both peaks A and A′ is approximately 90 meV at 300 K, much larger than the energy shift of the direct band edge, indicating an additional contribution from the large coupling between the bound exciton and optical phonons. Such a result is consistent with the analysis of temperature-dependent linewidth broadening of the peak for unconventional A exciton (see Sec. 5 of SI for more details).

Using optical spectroscopic measurements with the support of *ab initio* many-body calculations, we demonstrated the observation of an unconventional bright exciton in layered $SiP_2$. In contrast to those reported 1D and 2D excitons truly confined in CNTs and monolayer transition metal dichalcogenides, the bound excitonic states in layered $SiP_2$ exhibit hybrid low dimensionality due to the intrinsic 1D and 2D nature of the constituent electrons and holes, respectively. Interestingly, we envision that $SiP_2$ can host peculiar trion states, including a negatively charged trion (composed of two 1D-confined electrons and one 2D-confined hole) and a positively charged trion (composed of one 1D-confined electron and two 2D-confined holes). Once we couple layered $SiP_2$ to other vdWs semiconductors, such as monolayer $MoS_2$, to form heterostructures, the interfacial layer coupling can change the rotational symmetry of the semiconducting layers and could bring unprecedented new optical and optoelectronic functionalities via symmetry engineering at the heterointerfaces. Through the doping modulation of carrier polarity in $SiP_2$ or its heterostructures, rich excitonic physics with exotic dynamic behaviour can be realized in this material platform, such as interlayer excitons and Moiré excitons with tunable dimensionality. Furthermore, the interaction between this unconventional bound exciton and the optical phonon leads to an accompanying phonon sideband. Since a phonon and an exciton fall within the same energy range from zero to several hundred meV, we speculate that such many-body interactions may even lead to the emergence of novel elementary excitations beyond the Born–Oppenheimer limit in atomic 2D thin films or nanostructures of $SiP_2$. Our work will provide a new platform to further understand ex–ph coupling and other essential many-body physics and inspire follow-up studies and calculation method developments therein.

## Methods

**SiP$_2$ crystal growth with flux method**

Single crystalline samples were synthesized by using the Sn flux method[37]. Si, P, Gd and Sn were mixed at a Si:P:Gd:Sn ratio of 1:6:0.03:5 and sealed into evacuated quartz tubes. The mixture was slowly heated to 1100 °C to avoid bumping phosphorous and kept for 48 h. Subsequently, the sample was cooled to 400 °C in 140 h and then cooled to room temperature by switching off the electric furnace. The Sn flux was removed by using diluted HCl (aq). The obtained black-coloured crystals were then ultrasonicated in distilled water and ethanol to remove the residuals (such as phosphorous, adhered) on the crystal surface. This procedure was repeated until the water (or ethanol) became transparent enough after ultrasonication.

**Sample preparation for optical measurements and STEM–ADF measurements**

SiP$_2$ flakes with thicknesses from 5 nm to 200 nm were prepared by mechanical exfoliation onto SiO$_2$/Si wafers (300-nm-thick SiO$_2$ layer) or fused silica substrate. The thickness was identified by atomic force microscopy (AFM, integrated with WITec Alpha 300) after all optical measurements were finished. SiP$_2$ is stable in a nitrogen atmosphere and in vacuum and can gradually degrade when exposed to air within several hours. To avoid sample degradation, the whole sample preparation was processed in a glove box. Atomic-resolution STEM–ADF imaging was performed on an aberration-corrected ARM200F equipped with a cold field-emission gun operating at 80 kV. The STEM–ADF images were collected using a half-angle range from ~ 81 to 280 mrad. The convergence semi-angle of the probe was ~ 30 mrad.

**Optical measurements**

Optical measurements, including the PL spectra and RC and Raman spectra, were performed using a confocal Raman system (WITec Alpha 300). Thickness-dependent PL measurements (Sec. 3 of SI) were carried out at room temperature using a ×100 objective lens with an incident laser (laser power of 0.2 mW) focused to an ~ 1 μm spot. Nitrogen conditions were accomplished by protecting samples using continuous nitrogen gas flow. Low-temperature PL and Raman measurements were performed under vacuum conditions with samples installed in a cryostat (Cryo Instrument of America RC102–CFM Microscopy Cryostat) using a long working distance ×50 objective lens (laser power of 3 mW). For RC measurement, we recorded the subtracted reflectance of the sample normalized by the reflectance of the substrate, that is, $\text{RC} = 1 - \frac{R_{\text{sample}}}{R_{\text{sub}}}$, where $R_{\text{sample}}$ represents the reflectance of the SiP$_2$ sample on the silicon dioxide or quartz substrate and $R_{sub}$ represents the reflectance of the bare substrate.

**First-principles calculations**

First-principles density functional theory (DFT) calculations were performed by using the projector-augmented wave (PAW)[38,39] method implemented in the Vienna *ab initio* Simulation Package (VASP)[40]. The energy cut-off for the plane wave basis is set to 500 eV. To test the lattice constants to compare with the experimental value, we used the exchange-correlation functionals of generalized gradient approximation (GGA) with Perdew–Burke–Ernzerhof (PBE) type, local density approximation (LDA), and the PBE functional with vdWs corrections to fully relax the lattice structures. The vdWs interactions were included by using the methods proposed by Dion *et al.*[41] with the optB88-vdW functional. We found that lattice constants obtained from the method including the vdWs corrections are closest to the experimental values (see Secs. 10 and 11 of SI), which are used in the following phonon bands, *GW*, and *GW*–BSE calculations. During the lattice relaxations, the force convergence criterion was $10^{-3}$ eV/Å, and a 9×21×7 **k**-point mesh was sampled over the BZ. For the self-consistent electronic structure calculations, we set the energy convergence criterion to $10^{-6}$ eV and the **k**-point mesh to 11×25×9 over the whole BZ. The phonon spectrum was calculated by the PHONOPY package[42] in the framework of density functional perturbation theory (DFPT) with the finite-displacement approach, in which a $2\times4\times1$ supercell was employed.

Using the Vienna *Ab initio* Simulation Package[43], *GW* calculations[44] were performed using Kohn–Sham DFT wavefunctions (GGA–PBE) calculated on a 4×16×4 **k**-point mesh as the initial mean field. The dielectric response function used for the fully frequency-dependent eigenvalue-self-consistent *GW* calculation is summed over 1240 Kohn–Sham states (corresponding to a 100 eV cut-off). The frequency grid is divided into a dense part ranging from 0 eV to 13.75 eV and a coarse grid tail ranging from 13.75 eV to 178.78 eV. The grid sampling is nonuniform with 80 frequency grid points, resulting in step sizes ranging from 0.31 eV in the dense grid up to 46.35 eV in the tail[45]. We use multiple iterations in the *GW* calculation to update the eigenvalues of the Kohn–Sham states when calculating both Green's function *G* and the screened interaction *W* while keeping the initial Kohn–Sham wavefunctions unchanged. Full convergence was reached after five iterations. This procedure results in better agreement with the experimental results, as the standard $G_0W_0$ approach underestimates the bandgap by approximately 220 meV. The maximally localized Wannier functions obtained from the Wannier90 packages[46,47] were used to plot the *GW* quasiparticle band structure (Fig. 1e). In the construction of the Wannier functions, the *s* and *p* orbitals of both the Si and P atoms were used as initial trial wavefunctions. The *GW* quasiparticle energies and Kohn–Sham wavefunctions are used to construct the kernel of the BSE[48,49]. We employed the standard Tamm–Dancoff approximation and included 10 conduction bands and 10 valence bands during the calculation of the *GW*–BSE Hamiltonian.

**Calculation of the Spectral Function of the Phonon sidebands**

To model the spectral function of the phonon sidebands[50], we solved for the dressed polaron Green's function of the generalized Holstein Hamiltonian[51,52], $H = H_0 + V$, where $H_0 = \sum_{\mathbf{q}} \omega_{\mathbf{q}} b_{\mathbf{q}}^\dagger b_{\mathbf{q}} + \sum_{\mathbf{k}} \epsilon_{\mathbf{k}} c_{\mathbf{k}}^\dagger c_{\mathbf{k}}$ is the unperturbed single-particle Hamiltonian and $V = \sum_{\mathbf{k},\mathbf{q}} M_{\mathbf{k},\mathbf{q}} c_{\mathbf{k}+\mathbf{q}}^\dagger c_{\mathbf{k}} (b_{\mathbf{q}} + b_{-\mathbf{q}}^\dagger)$ is the interaction Hamiltonian. For the first term constituting $H_0$, $\mathbf{q}$ is the crystal momentum of the phonon, $b_{\mathbf{q}}^\dagger$ and $b_{\mathbf{q}}$ are the phonon creation and annihilation operators, and $\omega_{\mathbf{q}}$ (which shall be taken as a constant independent of $\mathbf{q}$) is the average phonon energy of the dominant optical branch responsible for ex–ph coupling. For the second term constituting $H_0$, $\mathbf{k}$ is the center-of-mass momentum of the exciton, $c_{\mathbf{k}}^\dagger$ and $c_{\mathbf{k}}$ are the exciton creation and annihilation operators, and $\epsilon_{\mathbf{k}}$ is its energy dispersion. We obtain these values from our *GW* calculations. The interaction Hamiltonian, $V$, represents the ex–ph interaction, with $M_{\mathbf{k},\mathbf{q}}$ being the ex–ph coupling matrix element. Since there is only one exciton in the model Hamiltonian, its solution is independent of the statistics of particle[1]. The same solution would be obtained for any fermion or boson, such as electrons (which is more common), as long as the particles are free to move. To construct the generalized Holstein Hamiltonian, we first calculate most of its parameters from first-principles calculations and finally fit the ex–ph coupling matrix elements, $M_{\mathbf{k},\mathbf{q}}$ (also to be taken as a constant), to the experimental results.

First, we note that although more than one phonon mode contributes to the renormalization of the band gap, the dominant contributing phonon modes fall within the energy range of 50 to 60 meV (Figs. 2f and S24). Since the phonon bands that project strongly onto the P$_B$–P$_B$ chain are relatively flat within the X–Γ–Z plane in reciprocal space (Fig. S23b), we assume the representative phonon band to have negligible phonon bandwidth, as in the Einstein model. Hence, we use the energy of a representative longitudinal optical (LO) phonon mode, $\omega_{\mathbf{q}} \equiv \omega_{\text{LO}} = 55$ meV, to model the ex–ph interaction, as obtained from the *ab initio* phonon calculations. We also assumed that the exciton has a free-particle dispersion of a periodic 1D chain, namely, $\epsilon_{\mathbf{k}} = -2t_{\text{ex}} \cos(ka)$, where $4t_{\text{ex}} = -80$ meV is the exciton bandwidth. The exciton hopping term, $t_{\text{ex}}$, was estimated using the hole bandwidth ($4t_{\text{h}} \approx -80$ meV) and the electron bandwidth ($4t_{\text{e}} \approx 640$ meV), which are calculated along the X–Γ–Z direction of the Brillouin zone from our *GW* calculations (Fig. S22), with the formula $t_{\text{ex}}^{-1} = t_{\text{h}}^{-1} + t_{\text{e}}^{-1}$. Finally, using the above parameters obtained from the first principles calculations, we only fit the ex–ph parameter, $M_{\mathbf{k},\mathbf{q}} \equiv M = 30$ meV, so that the spectral function of the calculated dressed Green's function reproduces the PL spectrum shown in the optical experiments (Fig. 2e).

In the calculation of the dressed interacting polaron Green's function, $G(\mathbf{k}, \omega)$, Dyson's identity, $[G(\mathbf{k}, \omega)]^{-1} = [G_0(\mathbf{k}, \omega)]^{-1} - \Sigma(\mathbf{k}, \omega)$ is used, where $G_0(\mathbf{k}, \omega)$ is the free-particle Green's function, given by $G_0(\mathbf{k}, \omega) = (\omega - \epsilon_{\mathbf{k}} + i\eta)^{-1}$ and $\Sigma(\mathbf{k}, \omega)$ is the ex–ph self-energy, which consists of an infinite sum of all proper self-energy diagrams. Written more explicitly, $G(\mathbf{k}, \omega)$ can be written as a continued fraction,

$$G(\mathbf{k},\omega)$$

$$= \cfrac{1}{G_0^{-1}(\mathbf{k},\omega) - \cfrac{M^2}{G_0^{-1}(\mathbf{k},\omega-\omega_0) - \cfrac{2M^2}{G_0^{-1}(\mathbf{k},\omega-2\omega_0) - \cfrac{3M^2}{G_0^{-1}(\mathbf{k},\omega-3\omega_0) - \cdots}}}}$$

$$= \frac{1}{G_0^{-1}(\mathbf{k},\omega) - \Sigma(\mathbf{k},\omega)},$$

such that $\Sigma(\mathbf{k},\omega)$ is the second term in the denominator given by

$$\Sigma(\mathbf{k},\omega) = \cfrac{M^2}{G_0^{-1}(\mathbf{k},\omega-\omega_0) - \cfrac{2M^2}{G_0^{-1}(\mathbf{k},\omega-2\omega_0) - \cfrac{3M^2}{G_0^{-1}(\mathbf{k},\omega-3\omega_0) - \cdots}}},$$

that when expanded in powers of $M^2$, reproduces the Feynman diagrams of each order[53-55]. In the calculation of the self-energy, we used the momentum-averaged (MA) noninteracting Green's function, as introduced by Berciu[54] and extended by Goodvin, Berciu and Sawatzky[55]. In this approximation, the momentum-dependent noninteracting Green's function, $G_0(\mathbf{k},\omega)$, in the expression of the self-energy, was replaced by its momentum average (MA), $\bar{G}_0(\omega)$, given by $\bar{G}_0(\omega) = \frac{1}{N_\mathbf{k}} \sum_\mathbf{k} G_0(\mathbf{k},\omega) = \int_{-\infty}^{\infty} d\epsilon \rho_0(\epsilon) G_0(\epsilon,\omega) = \frac{\text{sgn}(\omega)}{\sqrt{(\omega+i\eta)^2 - 4t_{xct}^2}}$, where $N_\mathbf{k}$ is the number of $\mathbf{k}$-points and $\rho_0(\epsilon)$ is the density of states. The momentum-averaged self-energy, $\Sigma_{MA}(\omega)$, is now momentum independent, and the interacting Green's function is now $G(\mathbf{k},\omega) = \frac{1}{G_0^{-1}(\mathbf{k},\omega) - \Sigma_{MA}(\omega)}$. Finally, the spectral function was given by the imaginary part of the interacting Green's function, the main peak of which is fitted to the excitation energy of excitonic state A as obtained from *GW*–BSE calculations.

## Data availability

The data that support the plots within this paper and other findings of this study are available from the corresponding authors upon reasonable request.

(2006).

55. Goodvin, G. L., Berciu, M. & Sawatzky, G. A. Green's function of the Holstein polaron. *Phys. Rev. B* **74**, 245104 (2006).## Acknowledgements

This research was supported by the National Natural Science Foundation of China (91750101, 21733001, 52072168, 51861145201), the National Key Basic Research Program of the Ministry of Science and Technology of China (2018YFA0306200), the Fundamental Research Funds for the Central Universities (021314380078, 021314380104, 021314380147) and Jiangsu Key Laboratory of Artificial Functional Materials. A.R. acknowledges the support from the European Research Council (ERC-2015-AdG-694097), Grupos Consolidados (IT1249-19), and the Max Planck-New York City Center for Non-Equilibrium Quantum Phenomena. The Flatiron Institute is a division of the Simons Foundation. P.Z.T. acknowledges the support from the Fundamental Research Funds for the Central Universities (ZG216S20A1) and the 111 Project (B17002). Part of the calculations were supported by the high-performance computing (HPC) resources at Beihang University. L.W. acknowledges funding by the Deutsche Forschungsgemeinschaft (DFG) under Germany's Excellence Strategy - Cluster of Excellence Advanced Imaging of Matter (AIM) EXC 2056 - 390715994 and by the Deutsche Forschungsgemeinschaft (DFG, German Research Foundation)–SFB-925–project 170620586. S.G.L. and C.S.O. acknowledge support by National Science Foundation Grant No. DMR-1926004 and National Science Foundation Grant No. OAC-2103991. X.X.Z. and A.T.S.W acknowledge support from MOE Tier 2 grant24

MOE2017-T2-2-139. X.X.Z. thanks the support from the Presidential Postdoctoral Fellowship, NTU, Singapore via grant 03INS000973C150. Y.F.L. acknowledge the support by Grant-in-Aid for Young Scientists (Japan Society for the Promotion of Science, JSPS) No. 21K14494. We acknowledge Chunfeng Zhang and Rui Wang for their support in pump-probe transient optical measurements and related data analysis.

**Author contributions**

L.Z., J.W.H., and L.W. equally contributed to this work. H.T.Y., P.Z.T., and A.R. conceived and designed the experiments and theoretical calculations. Y.F.L., H.G., and H.H. synthesized the bulk $SiP_2$ crystals. L.Z., C.R.Z., and M.T. performed sample fabrication. L.Z., J.W.H., C.Y.Q., and Z.Y.L. performed optical measurements and analysed the optical results. L.Z., C.Y.Q., and D.W. performed AFM measurements. L.Z. performed reflectance contrast simulations. X.X.Z., and A.T.S.W. performed STEM measurement. L.W., P.Z.T., and C.S.O. performed *ab initio* calculations and model simulations. L.W., P.Z.T., C.S.O., S.L., S.G.L., and A.R. analysed theoretical calculated results. L.Z., P.Z.T., H.T.Y., C.S.O., and A.R. wrote the manuscript with input from all authors. All authors contributed to the general discussion and revision of the manuscript.

**Competing interests**

The authors declare no competing interests.

**Correspondence and requests for materials** should be addressed to H.T.Y., P.Z.T. or A.R.

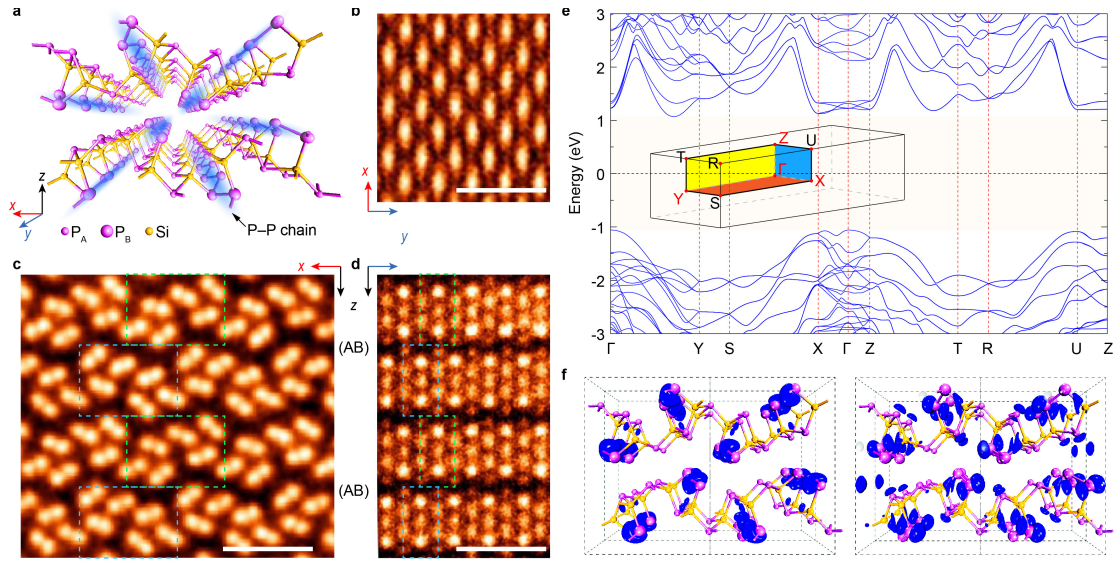

**Figure 1 Crystal structure and band structure of layered SiP$_2$**

**a**, Schematic layered structure of SiP$_2$. SiP$_2$ has *Pnma* phase (Group No. 62). Yellow spheres represent Si atoms, and pink spheres represent P$_A$ (small) and P$_B$ (large) atoms. The *xyz* coordinate system is defined according to the crystal structure, as shown in the bottom-left corner. The blue shadows highlight P$_B$–P$_B$ chains formed by the P$_B$ atoms along the *y* direction of the crystal lattice, which plays a critical role in generating quasi-1D electronic and excitonic states. **b**, Top view and **c-d**, cross-sectional STEM–ADF images of SiP$_2$. Green and cyan dashed rectangles in the STEM images represent the periodic lattice with ABAB stacking order of SiP$_2$ layers. Scale bars represent 1 nm for all three STEM images. **e**, Electronic band structure of bulk SiP$_2$ calculated from the *GW* method. The inset shows the first Brillouin zone of bulk SiP$_2$. It is a semiconductor with an indirect band gap of 2.14 eV. The valence band maximum (VBM) is at the Γ point, and the conduction band minimum (CBM) is located along the Γ–Y direction. The CBM state does not contribute to the formation of the A exciton owing to the large direct inter-band transition energies at this location. **f**, Charge density distribution of the conduction band edge (left panel) and valence band edge (right panel) in real space. The iso-surface of the plot is 0.02 e/Å$^3$.

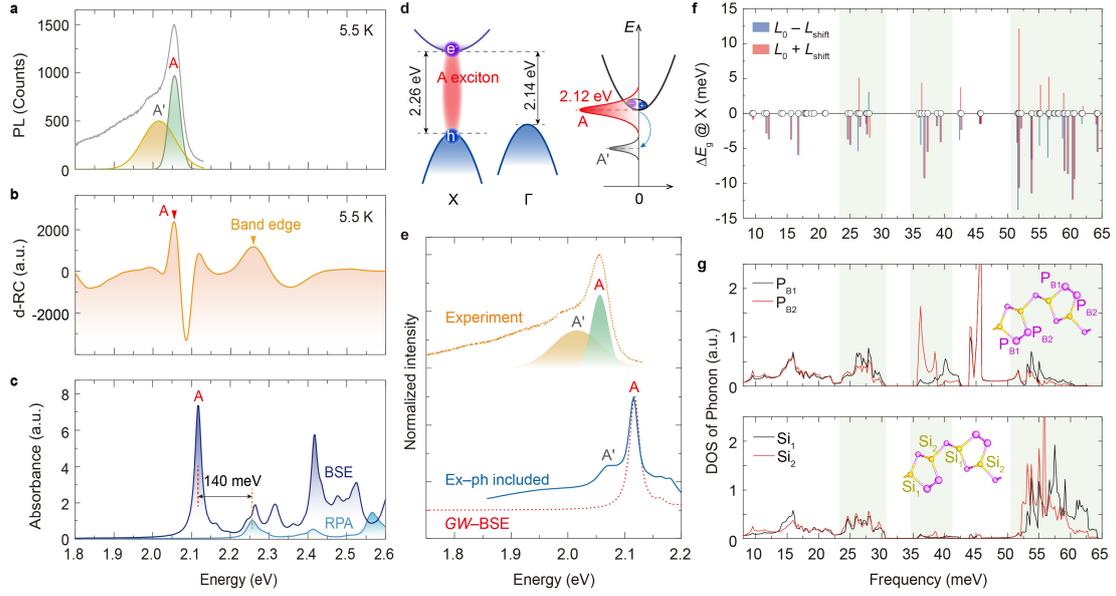

**Figure 2 PL, absorption and ab initio calculations of the A exciton and its sideband**

**a**, PL spectrum (gray solid line) measured at 5.5 K and its fitted results using two Gaussian peaks plus a background. The fitted peaks are assigned as A (green solid line) and A′ (yellow solid line). **b**, The 2$^{nd}$ derivative of the reflectance contrast (d-RC) spectrum (orange solid line) measured at 5.5 K. The arrows indicate the absorption peaks associated with the band edge (yellow) and the A exciton (red). **c**, Calculated absorption spectra of SiP$_2$ by using the *GW*–BSE (dark blue) and *GW*–RPA (cyan) methods. The yellow dashed line represents the band-edge transition, the red dashed line represents the A exciton, and the binding energy is 140 meV. **d**, Left panel shows schematic diagrams for excitons bound by the Coulomb interaction and electronic band structures for bulk SiP$_2$ with a direct band gap of 2.26 eV at the X point and an indirect band gap of 2.14 eV. Right panel shows the schematic diagram of the quasiparticle band for excitonic states, which includes the exciton peak A (2.12 eV) and the sideband A′.

**e**, Calculated absorption spectrum of SiP$_2$ with (blue solid line) and without (red dotted line) ex–ph interactions as well as the experimental PL spectrum (orange dashed line). The main A exciton peak of the blue line is obtained from *GW*–BSE calculations. The green and yellow shaded Gaussian peaks are A and A′, respectively, as defined in Fig. 2a. **f**, Energy shifts of the band gap at the X point $E_g^X$ individually induced by each optical phonon mode with momentum **q** = 0 at a temperature of zero. $L_0$ represents the lattice structure without displacement of phonon modes. $\pm L_{shift}$ stands for atomic displacements of phonon modes. The change in the band gap is estimated by averaging the energy shifts of the band gap $E_g^X$ with positive (red bar) and negative (dark blue bar) atomic displacements. Electronic structures and phonon eigenvalues are calculated by using the exchange-correlation functional with vdWs corrections. Black circles denote energies of the corresponding phonon modes. **g**, The phonon density of states for optical phonon modes, which is projected to the P$_B$ atoms in the embedded P$_B$–P$_B$ chains (top) and their neighboring Si atoms (bottom). In the insets, the P$_B$ atoms and their neighboring Si atoms are marked.

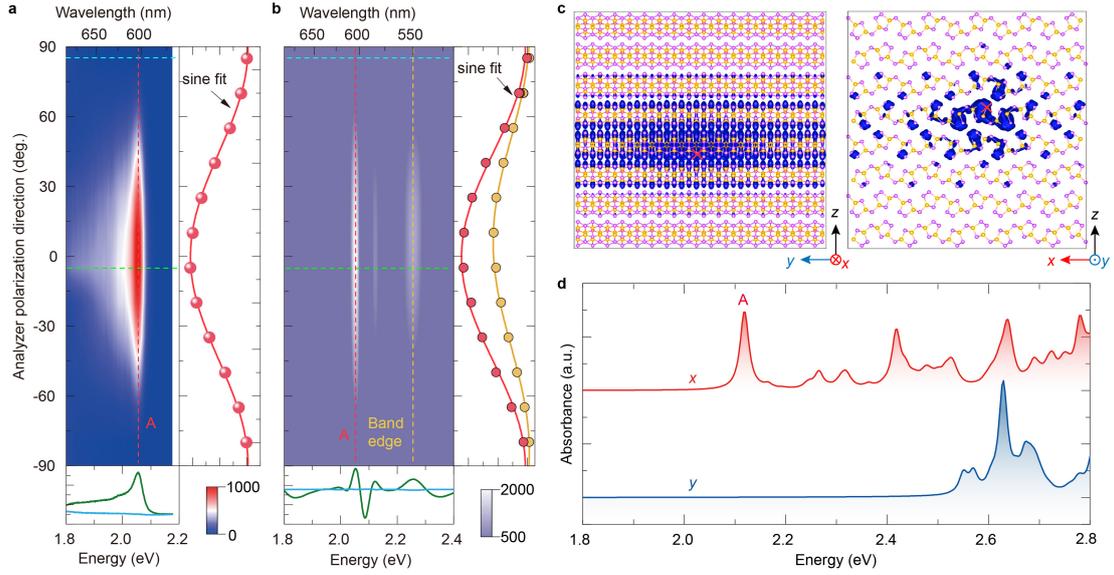

**Figure 3 Linear polarized nature of A exciton**

**a**, Contour plot of the PL intensity as a function of emission photon energy at different detection polarization angles $\theta$, which denotes the angle between the analyser polarization direction and the $x$ (defined in Fig. 1a) axis. Right panel: PL intensity and its sine fit versus detection angle along the red dashed vertical line at 2.06 eV. The character "A" indicates the position of the A exciton. Bottom panel: PL spectra with detection polarizations near 90° (cyan line) and 0° (green line). Colour bar indicates the PL intensity. **b**, Contour plot of the d-RC as a function of photon energy and detection polarization angle. Right panel: d-RC intensities (black circles filled by colours) and their sine fits (coloured solid lines) versus the detection angle along the yellow dashed vertical line at 2.26 eV and red dashed vertical line at 2.05 eV. The labels "A" and "Band edge" indicate the positions of the A exciton and band edge. Bottom panel: d-RC spectra with detection polarizations near 90° (cyan line) and 0° (green line). Colour bar stands for d-RC intensity. **c**, The modulus squared of the A exciton's wavefunction in real space distribution calculated from *GW*–BSE calculations. The red cross marks the position of the hole state. **d,** Simulated absorption spectra from the *GW*–BSE calculation along the *x* (red line) and *y* (dark blue line) directions. The absorption peak of the A exciton appears only when the polarization is along the *x* direction with the excitation laser incidents along the *z* axis.

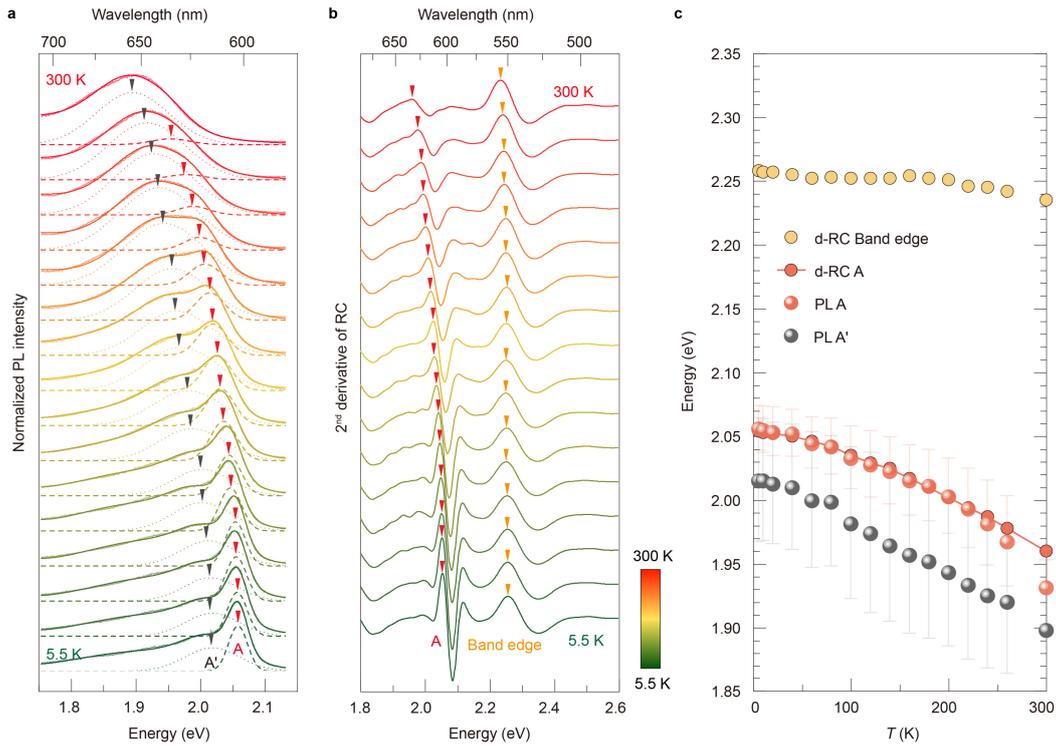

**Figure 4 Temperature-dependent spectra and energy evolution for exciton peak A and side peak A′ in SiP$_2$**

Temperature-dependent **a**, PL and **b**, d-RC spectra (2$^{nd}$ derivative of RC) from 5.5 K up to 300 K. The thick solid lines represent the spectra results, while the thin dashed and dotted lines represent fitting results of the PL A and A′ peaks. The solid triangle arrows highlight the redshifts of peak A (red), A′ (black) and the band edge (yellow) with increasing temperature. **c**, Temperature-dependent band edge (black circles filled by yellow), A exciton (black circles filled by red) energies extracted from d-RC spectra and A exciton (red sphere), A′ (black sphere) energies extracted by multipeak fitting of PL spectra. The error bar equals full width at half maximum.